\documentclass[sn-mathphys,Numbered]{sn-jnl}
\usepackage{graphicx}%
\usepackage{multirow}%
\usepackage{amsmath,amssymb,amsfonts}%
\usepackage{amsthm}%
\usepackage{mathrsfs}%
\usepackage[title]{appendix}%
\usepackage{xcolor}%
\usepackage{textcomp}%
\usepackage{manyfoot}%
\usepackage{booktabs}%
\usepackage{algorithm}%
\usepackage{algorithmicx}%
\usepackage{algpseudocode}%
\usepackage{subcaption}

\theoremstyle{thmstyletwo}%

\theoremstyle{thmstylethree}%

\begin{document}

\title[Article Title]{General Relativistic Approach to the Vis-viva Equation on Schwarzschild Metric}

\author*[1]{\fnm{Qi} \sur{Peng}}\email{pqi1003@gmail.com}

\author[1,2,3]{\fnm{Shuichiro} \sur{Yokoyama}}

\author[1,2]{\fnm{Kiyotomo} \sur{Ichiki}}

\affil*[1]{School of Science, Division of Particle and Astrophysical Science, Nagoya University, Chikusa-Ku, Nagoya, Aichi 464-8602, Japan}

\affil[2]{Kobayashi-Maskawa Institute for the Origin of Particles and the Universe, Nagoya University, Chikusa-Ku, Nagoya, Aichi 464-8602, Japan}

\affil[3]{Kavli IPMU (WPI), UTIAS, The University of Tokyo, 5-1-5 Kashiwanoha, Kashiwa, Chiba 277-8583, Japan}

\abstract{A modification to the vis-viva equation that accounts for general relativistic effects is introduced to enhance the accuracy of predictions of orbital motion and precession. The updated equation reduces to the traditional vis-viva equation under Newtonian conditions and is a more accurate tool for astrodynamics than the traditional equation. Preliminary simulation results demonstrate the application potential of the modified vis-viva equation for more complex n-body systems. Spherical symmetry is assumed in this approach; however, this limitation could be removed in future research. This study is a pivotal step toward bridging classical and relativistic mechanics and thus makes an important contribution to the field of celestial dynamics.}

\keywords{orbital motion, precession, vis-viva equation}

\maketitle

\section{Introduction}\label{sec1}

The concept of "vis viva," or "living force," has been a subject of scientific inquiry since the late 17th century. The German philosopher and mathematician Gottfried Wilhelm Leibniz first formalized this concept, laying the groundwork for what would later become the field of classical mechanics~\cite{smith2006, terrall2004}. Leibniz's work sparked a debate known as the "vis viva controversy," pitting him against the followers of René Descartes~\cite{iltis, jstor}. The controversy revolved around the correct formula for kinetic energy. Descartes and his followers advocated for $mv$ (mass times velocity) as the measure of "quantity of motion"\cite{Logsdon}. By contrast, Leibniz argued the formula $\frac{1}{2}mv^{2}$ better represented the "living force" of an object~\cite{mcdonough2021}. This formula is now universally accepted and forms the basis of our modern understanding of kinetic energy.

The 18th century saw further developments in this area, particularly by the Swiss mathematicians Johann and Jakob Bernoulli~\cite{terrall2004}. They extended Leibniz's concept to solve problems in celestial mechanics, a field that was gaining prominence because of works by Isaac Newton and others~\cite{jstor}. Bernoullis derived what is now known as the vis-viva equation from the conservation laws of classical mechanics. The equation is given for two interacting bodies as
\begin{eqnarray}
    v=\sqrt{G(M_1 + M_2) \left(\frac{2}{r}-\frac{1}{a}\right)},
\end{eqnarray}
where $G$ is the gravitational constant, $M_1$ and $M_2$ are the masses of the bodies, \(v\) is the relative velocity of the bodies, $r$ is the distance between the bodies, and \(a\) is the semimajor axis of the Keplerian orbit\cite{Logsdon}.

The classical vis-viva equation has been instrumental in understanding celestial mechanics; however, it has limitations. For instance, it does not account for relativistic effects that are important in systems such as Mercury orbiting the Sun. These limitations necessitate the development of a comprehensive model that incorporates general relativistic corrections.

In this note, we meet this need by introducing a general-relativistic extension to the vis-viva equation. This extension is particularly relevant for systems in which general-relativistic effects are significant, such as stars orbiting a supermassive black hole at the galactic center. Our preliminary results indicate that the modified equation converges to the classical Newtonian form under specific conditions and provides higher accuracy in numerical simulations than the traditional equation. The proposed approach also increases the precision between theoretical predictions of phenomena, such as the perihelion shift and numerical estimates.

\section{General-relativistic approach to the vis-viva equation}\label{sec2}

To derive a vis-viva equation for velocity, one starts from the geodesic equation. 
Here we assume the static and spherically symmetric background,
and based on Birkhoff's theorem the metric is given by
the Schwarzshild solution:
\begin{eqnarray}
    ds^{2}&=&g_{\mu\nu}x^{\mu}x^{\nu}\cr\cr
    &=&-\left(1-\frac{2GM}{c^{2}r}\right)dt^{2}+\left(1-\frac{2GM}{c^{2}r}\right)^{-1}dr^{2}+r^{2}(d\theta^{2
    }+\sin^{2}\theta d\phi^{2})~. \label{1}
\end{eqnarray}
Then, it yields the corresponding spatial component of the 4-velocity (on the 2-dimensional plane where $\theta =\frac{\pi}{2}$) as
\begin{eqnarray}
    \mathbf{u}^{2}=g_{ij}u^{i}u^{j}=\left(1-\frac{2GM}{c^{2}r}\right)^{-1}\left(\frac{dr}{d\tau}\right)^{2}+r^{2}\left(\frac{d\phi}{d\tau}\right)^{2}\label{2}.
\end{eqnarray}
From $g_{\mu\nu}u^{\mu}u^{\nu}=-c^{2}$ that leads to the energy conservation law~\cite{Schutz}, we obtain:
\begin{eqnarray}
    -\left(1-\frac{2GM}{c^{2}r}\right)^{-1}&E^{2}+\left(1-\frac{2GM}{c^{2}r}\right)^{-1}\left(\frac{dr}{d\tau}\right)^{2}+r^{2}\left(\frac{d\phi}{d\tau}\right)^{2}=-c^{2}\label{2.1}\\
    &\Rightarrow\mathbf{u}^{2}(r)=-c^{2}+\left(1-\frac{2GM}{c^{2}r}\right)^{-1}E^{2}.
\end{eqnarray}
Here, $E$ can be expressed in terms of the radial distance, i.e., $E=E(r)$. $E$ and the angular momentum of massive particles can be expressed within the context of general relativity as
\begin{eqnarray}
    \left\{ 
    \begin{array}{ll}
    E=p_{0}/m=-u_{0}~,   \\
    L=p_{\phi}/m=u_{\phi}~.
    \end{array}
    \right.
\end{eqnarray}
The term $\left(\frac{d\phi}{d\tau}\right)$ can then be rewritten as $\frac{L^{2}}{r^{2}}$.

In planetary motion, the shape of an orbit can be described as a precessing ellipse with a semimajor axis ($a$), semiminor axis ($b$), and eccentricity ($e$).  The two critical positions of a massive particle in the orbit are the \textit{Apogee} ($r_{a}$) and \textit{Perigee} ($r_{p}$), which can be written in terms of the aforementioned parameters as $r_{a}=a(1+e)$ and $r_{p}=a(1-e)$.
When a massive particle occupies either of these two critical positions, its radial velocity reduces to $0$. In this case,  Eq.~(\ref{2.1}) yields
\begin{eqnarray}
    E^{2}=\left(1-\frac{2GM}{c^{2}r}\right)\left(c^{2}+\frac{L^{2}}{r^{2}}\right)_{r=r_{a},r_{p}}\label{3}.
\end{eqnarray}
Through energy conservation, we obtain
\begin{eqnarray}
    \left(1-\frac{2GM}{c^{2}r_{a}}\right)\left(c^{2}+\frac{L^{2}}{r_{a}^{2}}\right)=\left(1-\frac{2GM}{c^{2}r_{p}}\right)\left(c^{2}+\frac{L^{2}}{r_{p}^{2}}\right).\label{3.1}
\end{eqnarray}
By making use of the parameters $a$ and $e$, which are related with $r_{a}$ and $r_{p}$ as $r_{a}+r_{p}=2a$ and $r_{a}r_{p}=a^{2}(1-e^{2})$~
\footnote{In the weak field limit, these parameters $a$ and $e$ respectively correspond to the semimajor axis and eccentricity.}
, the equation Eq.~(\ref{3.1}) can be rewritten as
\begin{eqnarray}
    L^{2}=GMac^{2}(1-e^{2})\left[1-\frac{GM}{ac^{2}}\cdot\frac{3+e^{2}}{1-e^{2}}\right].\label{4}
\end{eqnarray}
Analyzing the expression for the energy given in Eq.~(\ref{3}) considering the symmetry of the two critical positions at which $u(r)=0$ yields
\begin{eqnarray}
    E^{2}&=\frac{1}{2}\left[\left(1-\frac{r_{s}}{r_{a}}\right)\left(c^{2}+\frac{L^{2}}{r_{a}^{2}}\right)+\left(1-\frac{r_{s}}{r_{p}}\right)\left(c^{2}+\frac{L^{2}}{r_{p}^{2}}\right)\right]\nonumber\\
    &=c^{2}-\frac{r_{s}c^{2}}{a(1-e^{2})}+\frac{L^{2}(1+e^{2})}{a^{2}(1-e^{2})^{2}}\left[1-\frac{r_{s}}{a}\cdot\frac{1+3e^{2}}{(1-e^{2})(1+e^{2})}\right]~,
\end{eqnarray}
Where $r_{s}:=\frac{2GM}{c^{2}}$ is the Schwarzschild radius. Substituting the expression for $L$ given in Eq.~(\ref{4}) into the equation above, the energy is then expressed as
\begin{eqnarray}
    E^{2}=c^{2}-\frac{GM}{a}\cdot\frac{1-\frac{2r_{s}}{a(1-e^{2})}}{1-\frac{r_{s}(3+e^{2})}{2a(1-e^{2})}}.
\end{eqnarray}
Above all, the 3-velocity given by Eq.~(\ref{2}) (the vis-viva equation modified to include general relativistic effects) can be written as
\begin{eqnarray}
    \mathbf{u}^{2}=c^{2}\left\{\left(1-\frac{r_{s}}{r}\right)^{-1}\left[1-\frac{r_{s}}{2a}\cdot\frac{1-\frac{2r_{s}}{a(1-e^{2})}}{1-\frac{r_{s}(3+e^{2})}{2a(1-e^{2})}}\right]-1\right\}.\label{5}
\end{eqnarray}

In classical mechanics, the vis-viva equation is instrumental for calculating escape velocities, particularly for non-massive objects where general relativistic effects are negligible. However, in the context of massive celestial bodies, such as planets orbiting a star or solar systems orbiting galactic centers, the dynamics become significantly more complex. The modified vis-viva equation, incorporating general relativistic effects, is essential for accurately determining escape velocities in these scenarios.

Let us consider the case where the semimajor axis $a$ approaches infinity. In this limit, the modified vis-viva equation (Eq. \eqref{5}) simplifies to:
\begin{eqnarray}
    \mathbf{u}^{2}_{e} &=& \lim_{a \to \infty} c^{2} \left\{ \left(1 - \frac{r_{s}}{r} \right)^{-1} \left[1 - \frac{r_{s}}{2a} \cdot \frac{1-\frac{2r_{s}}{a(1-e^{2})}}{1-\frac{r_{s}(3+e^{2})}{2a(1-e^{2})}} \right] - 1 \right\} \\
    &=& c^{2} \left[ \left(1 - \frac{r_{s}}{r} \right)^{-1} - 1 \right] \\
    &=& c^{2}\left( \frac{r}{r_{s}} - 1 \right)^{-1}.
\end{eqnarray}
Thus, the escape velocity in a general-relativistic framework is given by:
\begin{eqnarray}
    \mathbf{u}_{e} &=& c\left( \frac{r}{r_{s}} - 1 \right)^{-1/2}.
\end{eqnarray}

This expression reveals that the escape velocity incorporates higher-order terms in comparison to the classical expression $v_{e}=\sqrt{\frac{2GM}{r}}$. This modification enhances the resolution for calculating escape velocities of massive objects, making the equation more applicable in various astrophysical contexts, including the study of event horizons and binary star systems.

\section{Justification of the modified vis-viva equation}
\subsection{Newtonian Limit of the General Vis-Viva Equation}
The general vis-viva equation is a cornerstone in gravitational dynamics for describing orbital motion. Equation ~(\ref{5}) can be rewritten by introducing 
 a modification term,  \( \alpha \), as follows

\begin{equation}
\alpha = \frac{1-\frac{2r_{s}}{a(1-e^{2})}}{1-\frac{r_{s}(3+e^{2})}{2a(1-e^{2})}} = 1 - \frac{(1 - e^{2})r_{s}}{2a(1-e^{2})-r_{s}(3+e^{2})}~.
\end{equation}

This modification term \( \alpha \) captures the relativistic effects that are absent in the classical model. In the Newtonian limit, where \( (1 - e^{2})r_{s} \ll 2a(1 - e^{2}) \), \( \alpha \) approaches $1$, consistent with the findings of Li-fang Li and Zhoujian Cao~\cite{li2023}. 
The expression for \( g_{rr} \) can be approximated in this limit as \( \left(1 - \frac{r_{s}}{r}\right)^{-1} \approx \left(1 + \frac{r_{s}}{r}\right) \),
whereby
Eq.~\eqref{5} becomes

\begin{equation}
\mathbf{u}^{2} = GM\left(\frac{2}{r} - \frac{1}{a}\right)~.
\end{equation}

This equation corresponds to the classical mechanical limit of the vis-viva equation with relativistic corrections. This form validates Eq.\ref{5} in the classical limit, in agreement with recent studies on galactic dynamics and dark matter, such as those by D. Astesiano et al.~\cite{astesiano2022} and A. Ryabushko et al.~\cite{ryabushko2022}.

\subsection{Numerical calculation of orbital motion and precession}

\begin{figure}[htbp]
    \centering
    \begin{subfigure}{0.49\textwidth}
        \centering
        \includegraphics[width=\textwidth]{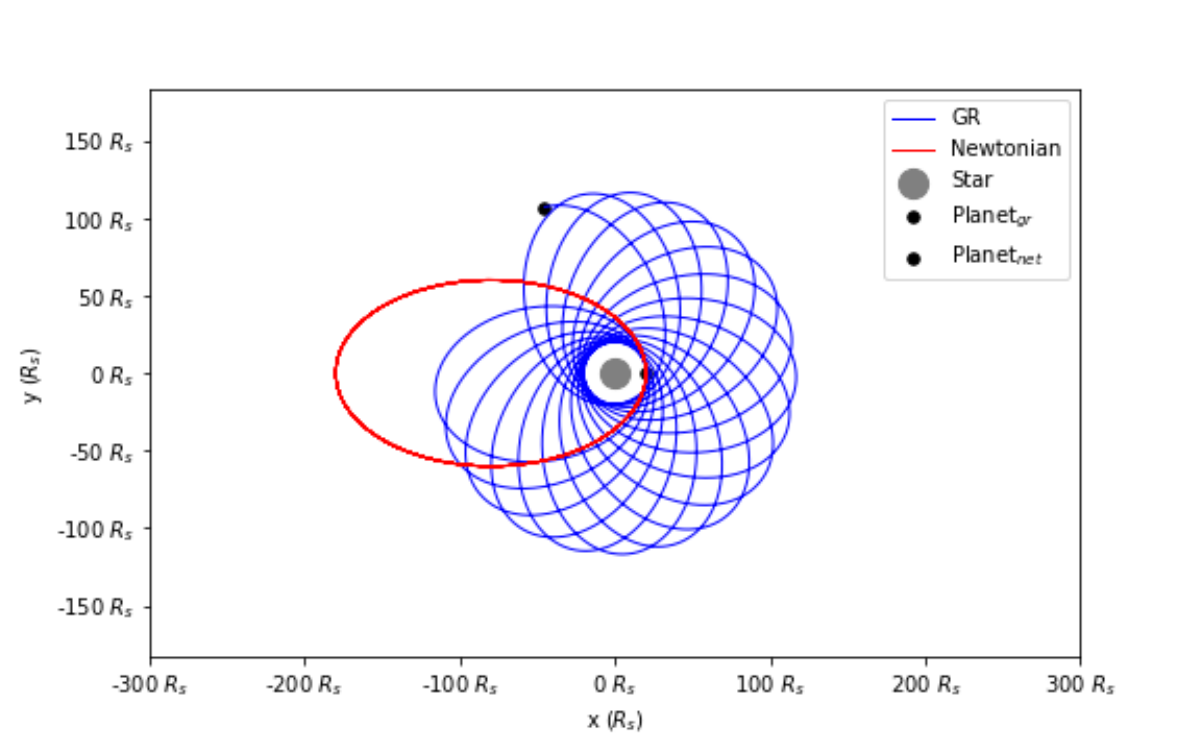}
        \caption{\small Precession calculated using the classical vis-viva equation}
    \end{subfigure}
    \vspace{5mm}
    \begin{subfigure}{0.49\textwidth}
        \centering
        \includegraphics[width=\textwidth]{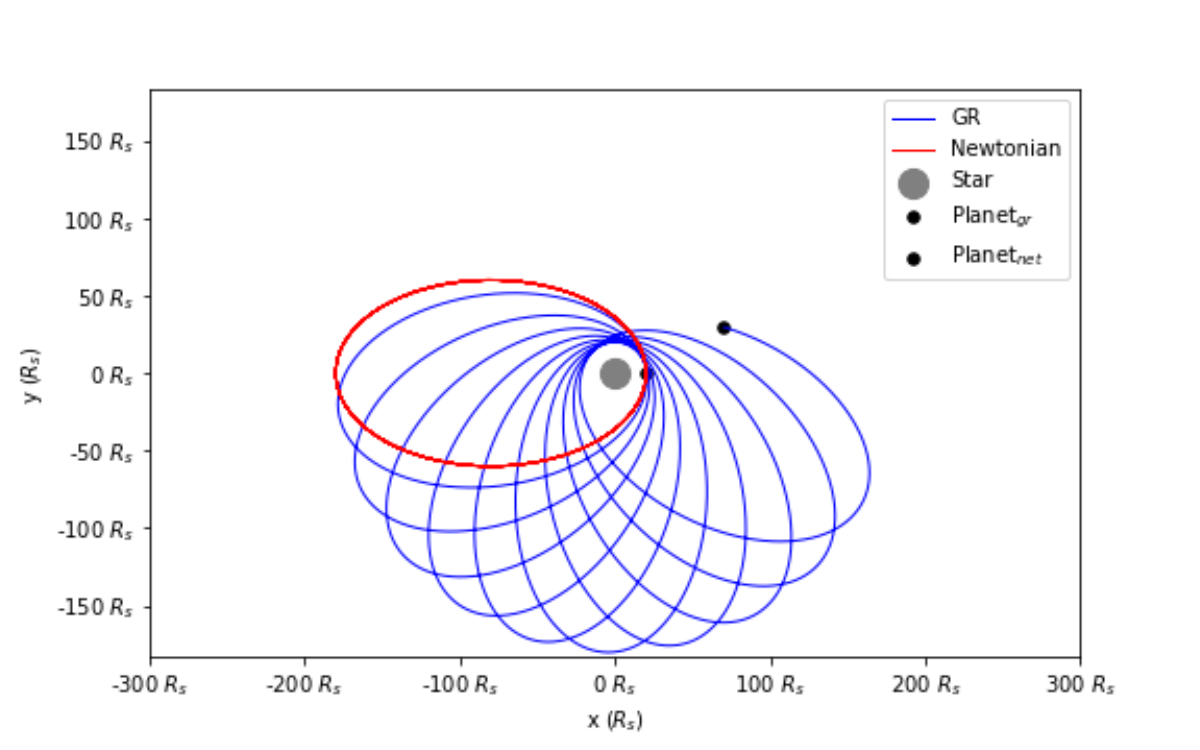}
        \caption{\small Precession calculated using the modified vis-viva equation}
    \end{subfigure}
    \caption{\small{Simulated precession of a planet, starting at the position of orbit perigee.}}
    \label{pic1}
\end{figure}

To demonstrate the importance of the general-relativistic extension of the vis-viva equation, we numerically calculate the orbital motion using the Shwarzschild metric. As an example, we consider a case where 
$a=100R_{s}$ and $e=0.8$ where $R_s$ is the Schwarzschild radius. The results are shown in Fig.~\ref{pic1}. 
In the calculation, the test particle is considered to be at the perigee of the orbit to set the initial conditions for the radial distance and velocity (momentum) in calculation, In Fig.~\ref{pic1}, the left panel shows the results obtained using the classical vis-viva equation, and the right panel shows the results using the vis-viva equation modified to include the general-relativistic effects. 
For reference, we show the Kepler elliptic motion as red lines for the same values of the orbital parameters. 
The result obtained using  the modified vis-viva equation with the specified initial condition correctly represents the precession of the perihelion,
whereas the semimajor axis of the orbit calculated using the classical vis-viva equation deviates from the actual orbit from the start of the simulation.

To highlight the differences between the results obtained using the two equations, we quantitatively evaluate the degree of the perihelion shift against the exact analytic solution. The exact solution for the degree of the perihelion shift (from one perihelion to the next) can be expressed in terms of the orbital parameters as
\begin{eqnarray}
    \Delta\phi =2\pi\left({1-\frac{6GM}{c^{2}a(1- e^{2})}}\right)^{-1/2} - 2 \pi,
\end{eqnarray}
and the results are summarized in Table~\ref{table1}.
In the Table, a ``theoretical estimate" is obtained using the exact analytic solution presented above,
and ``degree shift (classical/modified)" is the numerical result.
\begin{table*}[h]
\centering
\begin{tabular}{|c|c|c|}
\hline
steps interval & $\Delta\phi$[radians] & Accuracy (Extreme)  \\
\hline
Theoretical estimate & $0.2794$  & -\\
Degree shift (classical) & $0.2962$ & $93.98$\% \\
Degree shift (modified) & $0.2801$ & $99.75$\% \\
\hline
\end{tabular}
\caption{\small Comparison of the perihelion shifts numerically calculated using the classical and the modified vis-viva equations using fictitious model parameters.}
\label{table1}
\end{table*}

From the Table, one can see that the modified vis-viva equation has successfully raised the accuracy of perihelion shift with theoretical value, which justifies Eq.~(\ref{5}) in one aspect.

In summary, compared with the classical vis-viva equation, the modified one with general relativistic effects given by Eq.~(\ref{5}) has been demonstrated to provide a more accurate and precise estimate of the orbital motion of an astrophysical object
under relatively strong gravity, such as the stellar orbital motion around the massive black holes at the center of our Milky Way.

\section{Conclusion}
The accurate prediction of orbital motion holds paramount importance in astrophysics and celestial mechanics. While the classical vis-viva equation is foundational in these fields, its limitations become apparent in scenarios where general relativistic effects are prominent. To address this, our study has developed a general relativistic modification of the vis-viva equation, enhancing the precision in predicting orbital motion, particularly in regions under strong gravitational influence.

Our modified equation integrates seamlessly with the principles of general relativity, thereby proving more effective than its classical counterpart in predicting phenomena such as the perihelion shift of celestial bodies. This is especially relevant in systems where general relativistic effects are significant, such as in the dynamics of stars orbiting supermassive black holes at galactic centers. Notably, the equation converges to its classical form in the Newtonian limit, demonstrating both its robustness and versatility.

One of the compelling future applications of this modified vis-viva equation lies in the realm of N-body systems, akin to its classical usage but with a relativistic perspective. The extension to N-body systems would allow for a more comprehensive understanding of complex celestial dynamics, particularly in dense astrophysical environments like star clusters or galactic nuclei where relativistic effects cannot be ignored.

In addition to N-body systems, the modified vis-viva equation holds potential for application in binary systems of black holes under the post-Newtonian approximation. This approach would facilitate a more nuanced understanding of the orbital mechanics in binary black hole systems, contributing to the burgeoning field of gravitational wave astronomy. Such an application would not only validate the equation in a new context but also offer a tool for predicting the orbital evolution in these extreme environments.

Furthermore, considering black holes, extending the vis-viva equation to accommodate the Kerr metric presents an exciting avenue for future research. This adaptation would enable the exploration of orbital dynamics around rotating black holes, a scenario where the classical vis-viva equation is markedly insufficient. The Kerr-metric extension would significantly enhance our understanding of accretion dynamics and jet formation mechanisms around such astrophysical objects.

In conclusion, the modified vis-viva equation, as developed in this study, not only augments our current understanding of orbital dynamics in strong gravitational fields but also opens up multiple promising pathways for future research. These explorations could significantly contribute to our theoretical and practical understanding of various astrophysical phenomena, from the dynamics of binary black holes to the intricate dance of stars around supermassive black holes in galactic cores.

\bmhead{Acknowledgments}

SY is supported in part by the JSPS KAKENHI Grant
Number JP20K03968 and JP 23H00108.
KI is supported by the JSPS grant number 21H04467, JST FOREST Program JPMJFR20352935, and by JSPS Core-to-Core Program (grant number: JPJSCCA20200002, JPJSCCA20200003).

\bibliography{Reference}

\end{document}